\documentclass[12pt]{iopart}

\usepackage{graphicx}  
\begin{document}

\title[]{Experimental test of the Pauli Exclusion Principle}

\author{A S Barabash}

\address{Institute of Theoretical and Experimental Physics, 
B. Cheremushkinskaya 25, 117218 Moscow, Russia}

\ead{barabash@itep.ru}
\begin{abstract}

A short review is given of three experimental works on tests of the Pauli Exclusion Principle (PEP)
in which the author has been involved during the last 10 years. In the first work a 
search for anomalous carbon atoms was done and a limit on the existence of such atoms was 
determined, $^{12}\tilde{\mathrm C}$/$^{12}$C $< 2.5\times10^{-12}$. In the second work PEP was tested with the NEMO-2 
detector and the limits on the violation of PEP for p-shell nucleons in $^{12}$C were obtained. 
Specifically, transitions to the fully occupied $1s_{1/2}$-shell yielded a limit of $4.2\times10^{24}$ y 
for the process with the emission of a $\gamma$-quantum. Similarly limits of $3.1\times10^{24}$ y for $\beta^-$ and 
$2.6\times10^{24}$ y for $\beta^+$ Pauli-forbidded transition of 
$^{12}$C $\to$ $^{12}\tilde{\mathrm N}$($^{12}\tilde{\mathrm B}$) are reported. In the 
third work it was assumed that PEP is violated for neutrinos, and thus, neutrinos obey 
at least partly the Bose-Einstein statistics. Consequences of the violation of the exclusion 
principle for double beta decays were considered. This violation strongly changes the 
rates of the decays and modifies the energy and angular distributions of the emitted 
electrons. It was shown that pure bosonic neutrinos are excluded by the present experimental 
data. In the case of partly bosonic neutrinos the analysis of the existing data allows one to 
put an upper bound for $sin^2\chi < 0.6$. The sensitivity of future measurements is also evaluated.

\end{abstract}

\maketitle

\section{Introduction}

The Exclusion Principle is one of the most 
fundamental laws of nature. It was 
formulated by W.~Pauli in 1925 \cite{Pauli25} to explain the regularities 
of the Periodic Table of elements and the characteristic features of atomic
 spectra. In modern Quantum Field Theory (QFT) the Exclusion Principle appears 
automatically from the nature of identical particles and the 
anti-commutativity of the fermion creation (annihilation) operators. 
It postulates that in a system of identical fermions, two or more particles 
cannot occupy the same state.

The discovery in 1956 of parity non-conservation in $\beta$-decay 
\cite{Wu57} showed
for the first time that "fundamental laws" can be violated. The
violation of CP invariance was then discovered in 1964
\cite{Christinsen64}. As a 
result, all conservation laws began to be tested. Some of them, for
example, the non-conservation of leptonic and barionic quantum numbers
can be explained in the framework of models 
satisfying all the principles of standard QFT.
Others, such as the non-conservation of the electric charge 
\cite{Okun78,Ignatiev79,Voloshin78,Mohapatra92}, CPT-violation
\cite{Ellis97,Kostelecky97}, and Lorentz-invariance violation
\cite{Nielsen78,Kostelecky89}, require 
a global reconstruction of modern theoretical physics to create 
self-consistent models.

In more recent publications 
\cite{Ignatiev87,Okun87,Greenberg87,Greenberg89,GreenbergM89,Okun89}
some attempts 
were made to introduce into the theory a small violation of PEP,
but they have not been successful.
PEP is at the heart of the QFT and its violation, even if very small, leads to 
the appearance of states with negative norma (negative probability)
\cite{Govorkov89,Govorkov90}. 
Thus there is no answer to the question, "~What 
is the accuracy of PEP?". 
The reason for this is that there is no real self-consistent and
non-contradictory model, with small PEP violation.
Indeed any model with PEP violation
must be beyond the standard QFT. It was L.B.~Okun \cite{Okun89,Okun87}, 
who said," That exceptional place occupied by the Pauli principle
in modern physics does not imply that it does not need further painstaking 
experimental tests. Quite the opposite: the fundamental character of this 
principle generates special interest to its quantitative testing throughout
the Mendeleev Table".

Experimental searches of the effects of the 
Pauli principle violation via electrons (see, for example,
 \cite{BAR98,NOV90,JAV00,BART06,CUR08} and review \cite{IGN06})  
and nucleons (see, for example, \cite{ARN99,Logan79,Kekez90,BAR99,BOR04,BER97}) 
which have given negative results,  
leading to extremely strong bounds on the magnitude of 
the violation.  

In this paper results obtained in Ref. \cite{BAR98}, \cite{ARN99}, and \cite{BAR07} 
are presented.

\section{Search for anamalous carbon atoms is sought as evidence of the violation of PEP during the period of 
nucleosynthesis \cite{BAR98}}

This paper addresses a search for anomalous ("non-Paulian") atoms. Such atoms could be 
of cosmological origin, if not all $10^{80}$ electrons in the universe are antisymmetrized or if 
spontaneous transitions of ordinary atoms into "non-Paulian" atoms are nevertheless possible. 
The chemical properties of atoms with three electrons per 1s shell must be similar to the 
properties of their "lower-order" neighbors in the periodic table (for example, "non-Paulian" 
carbon would be similar to boron).

In 1989 Novikov and Pomansky proposed a check of PEP by searching for anomalous atoms 
arising from the periods of nucleosynthesis \cite{NOV89}. If PEP is violated, 
then every substance containing elements with the atomic number Z contains an admixture of anomalous 
atoms of the element with the atomic number (Z + 1), since these anomalous atoms have the same 
chemical properties as the element with the atomic number Z. The concentration of anomalous atoms 
in a substance is high in the case that the cosmic abundance of the parent element (Z + 1) is high, 
while that of the element Z is low. If the formation of "non-Paulian" atoms occurred as a result 
of a spontaneous transition of an outer electron into an inner shell, then the concentration 
of anomalous atoms in the material will be 

\begin{equation}
C = t\times P(Z+1)/\tau\times P(Z),
\end{equation}

where t is the average time which has passed since the moment when the anomalous atom formed 
to the end of presolar system formation ($\sim 4.5\times10^{9}$ y \cite{NOV89}); $\tau$ is the lifetime of an 
atomic electron with respect to the violation of PEP; and, P(Z + 1) and P(Z) are the cosmic 
abundances of elements with atomic numbers (Z + 1) and Z. 

In Ref. \cite{NOV89} two pairs of atoms were proposed as the most promising for the investigation:
boron-carbon and fluorine-neon, with the ratios $P(Z + 1)/P(Z) = 2.18\times10^6$ and 650, 
respectively. An experimental search for anomalous neon atoms in fluorine and argon atoms in 
chlorine has been performed \cite{NOV90} (the results obtained by accelerator mass spectrometry are 
$\tau > 2\times10^{30}$ y and $\tau > 4\times10^{27}$ y), while the boron-carbon pair 
remained unstudied. 

The work \cite{BAR98} was devoted to the search for anomalous carbon atoms in boron. 
Recall that boron exists in the form of two stable isotopes, $^{10}$B ($\sim$ 19\%) and 
$^{11}$B ($\sim$ 81\%). Carbon likewise consists of two stable isotopes, $^{12}$C ($\sim$ 99\%) 
and $^{13}$C ($\sim$ 1\%).The anomalous atom 
$^{12}\tilde{\mathrm C}$ contains three K-shell electrons and therefore behaves chemically like a boron 
atom. Such anomalous atoms should be concentrated in boron and its compounds in the process of 
evolution. In order to observe them a search must be conducted for anomalous $^{12}$C nuclei 
in boron or for boron atoms with a nuclear mass of 12.

In the present work the first possibility was investigated. In discussing possible 
experiments in search of anomalous carbon atoms in boron, Novikov and Pomansky proposed mass 
spectrometry and even estimated the sensitivity of experiments of this kind. However, their 
estimate is much too high, because of a large admixture of "ordinary" carbon (at the 0.1\% level)
is always present in boron samples and in the residual atmosphere of the mass spectrometer.

The idea of this experiment was to remove carbon atoms from a boron sample and then to measure 
the content of carbon nuclei in it. The search for anomalous carbon atoms was conducted by 
$\gamma$-activation analysis of different boron samples. Boron is ideal for the 
$\gamma$-activation analysis, since irradiation with $\gamma$-quanta does not produce 
radioactive isotopes and the activity is thereby determined by 
the impurities.

The experiment was performed on the microtron at the Institute of Physics Problems of the Russian 
Academy of Sciences. The boron samples for analysis consisted of amorphous boron (powder, obtained from 
BCl$_3$ gas), $\alpha$- and $\beta$-rhombohedral boron, boron whiskers on tungsten, decaborane 
(B$_{10}$H$_{14}$), and boron grown by zone melting. Experiments showed that the purest 
boron was obtained by zone melting, and boron from this source was used for the most conclusive measurements. 
In these samples the initial content of carbon atoms was equal to approximately $1.5\times 10^{-3}$ g/g. 
For this reason, it is impossible to search for anomalous carbon atoms in boron by the method 
of nondestructive $\gamma$-spectrometry on irradiated samples, and another special method was 
developed for removing impurities from the boron.

A boron sample with a mass of $\sim$ 0.1 g was irradiated with $\gamma$-quantum bremsstraglung from
28-MeV electrons. If $^{12}$C nuclei are present in the boron, then the 
reaction $^{12}$C($\gamma$,n)$^{11}$C 
should occur, i.e., radioactive $^{11}$C nuclei with a half-life of 20.34 min are formed.
In 99\% of the cases $^{11}$C undergoes $\beta^+$ decay and can therefore be 
detected by the $\gamma-\gamma$ coincidence method (the detection of two 511 keV annihilation
$\gamma$-rays, which are emitted after the positrons stop and are annihilated in the material).

After irradiation, the boron sample was oxidized in a KNO$_3$-KOH (3:1) melt for about 10 min at 600 $^o$C.
In the process of oxidative melting in an alkaline melt, the carbon-containing impurities in 
boron (a solid solution of carbon, i.e., boron carbide B$_4$C) are oxidized to CO$_2$, CO, KHCO$_3$, 
and K$_2$CO$_3$, while boron is oxidized to B$_2$O$_3$. The cooled melt was then dissolved in 
water containing nitric acid to convert the boron anhydride into boric acid H$_3$BO$_3$. Next, 
the solution was heated to the boiling point. In this process the potassium bicarbonates and 
carbonates decomposed, and carbon dioxide gas was released. The carbon dioxide was removed from the 
boric acid solution by a flow of air. Experiments on measuring the activity of the released $^{11}$CO$_2$
showed that in 25 min such a procedure removes more than 99\% of the carbon from the solution. After 
the CO$_2$ was driven off, the change in the activity of $^{11}$C nuclei remaining in the boric 
solution was measured as a function of time. Graphite was irradiated simultaneously with the sample. 
Measurements of the activity of the graphite made it possible to calculate the mass of the remaining carbon.

The plot displayed in Fig. 1 shows the results of one of these measurements. The residual activity 
is due essentially to the background which remains at approximately 40 decays per 100 s and only
a slight elevation in activity at the beginning of the measurements can be attributed to radioactive 
carbon nuclei with 20.34-min half-life. The inset in the figure displays on a large scale the result 
of the measurements of the activity of the solution and the result of the mathematical analysis of the data 
using the function $a+b\times exp(-t/T_{1/2}$), where $a$ is the constant background and $T_{1/2}$ = 1224 s 
is the half-life of $^{11}$C.

\begin{figure}
\begin{center}
\resizebox{0.75\textwidth}{!}{\includegraphics{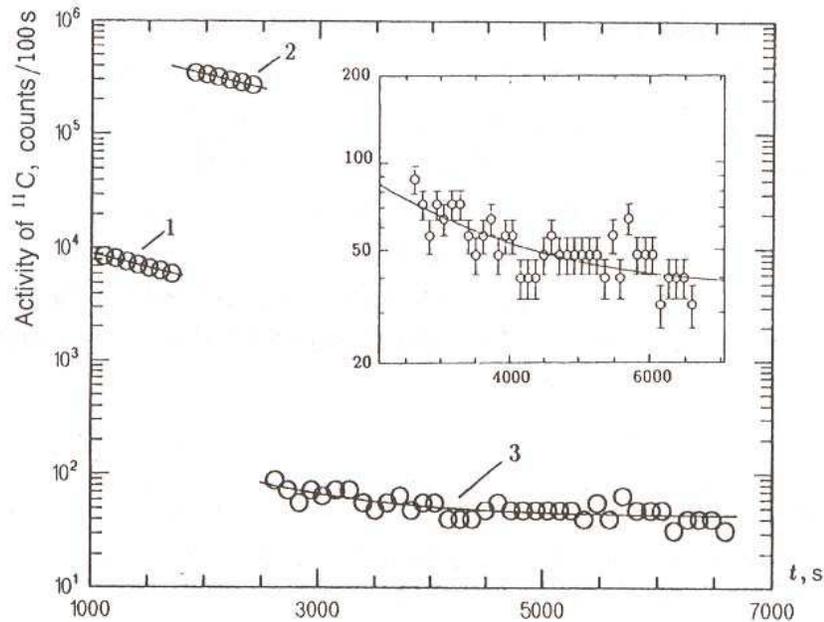}}
\label{fig:figfig_C_B}
\caption{Time dependence of the positron activity of carbon separated from a 100 mg boron sample 
after irradiation (1), a 6 mg sample irradiated together with the experimental sample (2), and 
a solution after chemical separation of carbon atoms (3). Inset: the detected positron activity 
of the solution. The solid line is a fit of the data assuming that only $^{11}C$ radionuclids and the background are present. (From Ref. \cite{BAR98}).}
\end{center}
\end{figure}

A series of measurements of this kind established that the concentration of anomalous carbon atoms does 
not exceed $5\times 10^{-6}$ g/g. The point is that one cannot rule out the possibility that the observed activity 
of $^{11}$C nuclei is due to the residual content of "normal" carbon in the boron sample. For this reason,
it can only be asserted that the concentration of anomalous $^{12}\tilde{\mathrm C}$ atoms 
in boron is $< 5\times10^{-6}$ g/g. 
Since 
the cosmic abundance of carbon is $2\times10^6$ times greater than that of boron, this means that the relative
concentration of anomalous carbon atoms in carbon corresponds to 
$^{12}\tilde{\mathrm C}$/$^{12}$C $< 2.5\times10^{-12}$. 
Using this limit 
obtained, one finds from Eq.(1) that the lifetime of electrons in a carbon atom relative to 
the violation of the Pauli principle is $\tau > 2\times10^{21}$ y.

In closing, note that the sensitivity of the $\gamma$-activation method, which was used to determine 
the $^{12}$C concentration in boron, would in principle permit increasing the limit of the determination of $\tau$ 
by an 
order of magnitude if one could obtains a boron sample with a low content of "normal" carbon impurity or
if the carbon could be effectively removed from the boron sample after irradiation.

\section{Testing PEP with the NEMO-2 detector \cite{ARN99}}

\subsection{NEMO-2 detector}

The NEMO-2 detector \cite{NEMO-2} was designed for double beta decay
 studies  and  operated in the Fr\'ejus Underground Laboratory (4800 m w.e.)
 from 1991 to 1997.  During this period, the
two neutrino double beta decays of $^{100}$Mo \cite{NEMO2-Mo}, 
$^{116}$Cd \cite{NEMO2-Cd}, $^{82}$Se \cite{NEMO2-Se} and 
$^{96}$Zr \cite{NEMO2-Zr}
were investigated in detail through the measurements of the summed electron
energy spectra, angular distributions and single electron spectra.  

The NEMO-2 detector (Fig. 2) consisted of a 1$\mbox{m}^3$
tracking volume filled with a mixture of helium gas and 4\% ethyl alcohol.
Vertically  bisecting the detector was the plane of the source foil under study
(1m $\times$ 1m). Tracking was accomplished with long, open Geiger cells
with an octagonal cross section defined by
100~$\mu$m nickel wires.
On each side of the source foil there were 10 planes of 32 cells
which alternated between vertical and horizontal orientations.

\begin{figure*}
\resizebox{0.75\textwidth}{!}{\includegraphics{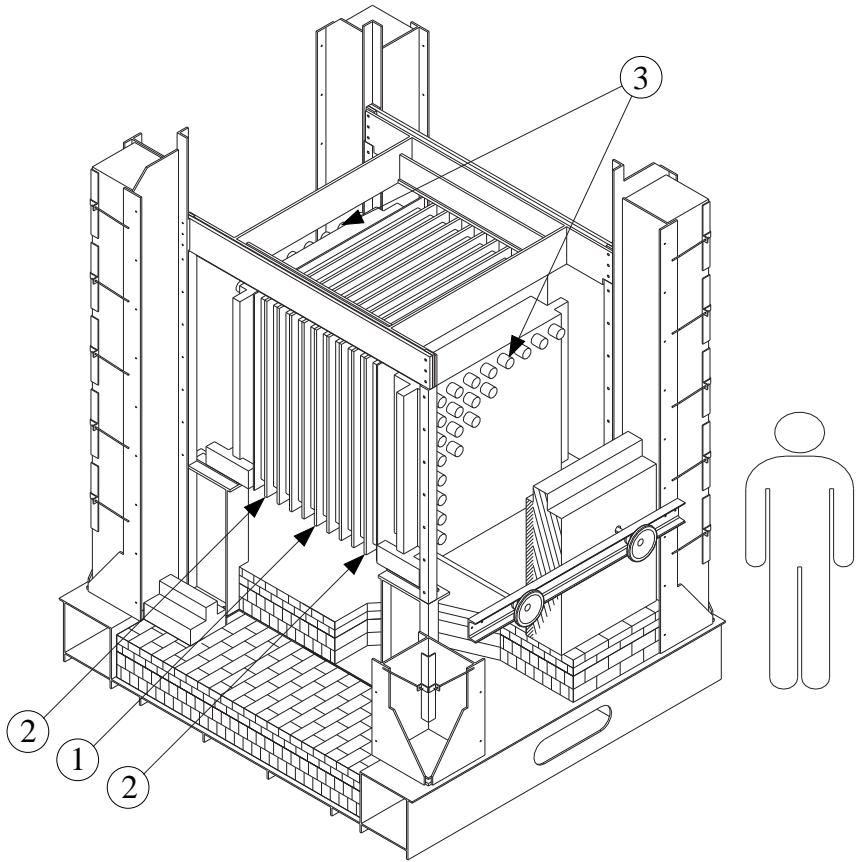}}
\label{fig:fignemo2d}
\caption{{The NEMO-2 detector without shielding.
(1) Central frame with the source plane was capable of supporting plural
    source foils.
(2) The tracking device of 10 frames, each consisting of two perpendicular
    planes of 32 Geiger cells.
(3) Two scintillator arrays each consisting of 5 by 5 counters for a calorimeter.
 In the earlier experiment with molybdenum sources \cite{NEMO2-Mo} the
 scintillator arrays were 8 by 8 counters as depicted here.}
}
\end{figure*}

A calorimeter made of scintillators covered two vertical opposing
sides of the tracking volume.
It consisted of
 two planes of  25 scintillators
(19 cm$\times$19 cm$\times$10 cm), combined with low radioactivity
photomultipliers tubes (PMT).

The tracking volume and scintillators were surrounded by
a lead (5~cm) and iron (20~cm) shield for measurements with
$^{100}$Mo and $^{116}$Cd. 
The same shield was used in the experiment with $^{82}$Se
and $^{96}$Zr foils for 6222.6 h
with the Zr foils placed at the central part of the source plane.
The lead was then
placed outside the iron for 1784.5 h. Next, the lead was removed 
for 536 h. At the end, 15 cm of paraffin was installed outside of
the iron for the final 2162.8 h.  

Details of the performance and parameters are described elsewhere
\cite{NEMO-2}, while the most salient
characteristics are outlined briefly here. Three-dimensional 
measurements of charged particle tracks are provided
by the array of Geiger cells.
The transverse position is given by the drift time, and the longitudinal
position is given 
by the plasma propagation times. The transverse resolution is
500~$\mu$m and the longitudinal resolution is 4.7~mm.
The calorimeter's energy resolution (FWHM) is 18\% at 1~MeV with a
time resolution of 275~ps (550~ps at 0.2~MeV). Scintillation counters  
measured the energy of an individual electron
in the interval from 50 keV to 4 MeV. 
Electrons with energies near or above $4$ MeV 
may fall into the "saturation" regime of the counters.
More specifically, $E_{sat}$ is different for each counter and varies
from 4 to 7 MeV. The only information for events in this 
 regime is that they deposit an energy higher than $E_{sat}$.
A laser and fiber optics device is used to check the stability of the
scintillation detectors.

A trigger requiring one or two scintillation counters and four Geiger frames
normally run at a rate of 0.01-0.04~Hz
depending on the radon levels in the laboratory. This trigger
rate is too low for an efficient calibration survey of the experiment, so a second trigger
requiring only one counter with an energy greater than 1.3~MeV was added.

An electron is defined by a track linking the source foil and one scintillator. 
The maximum scattering angle along the track has to be less
than $\mbox{20}^{\circ}$ to reject hard scattering situations.
A photon is recognized as one or two adjacent-fired 
scintillators, without an associated particle track. For photons
and electrons, an energy deposited greater than 200~keV is required in
order to obtain  sufficiently good time resolution for time-of-flight 
analysis of events. 
The two-electron events are defined by two tracks which have
 a common vertex in the source foil and are associated with two fired 
scintillators. 
A more detailed description can be found in the following references 
\cite{NEMO-2,NEMO2-Mo,NEMO2-Cd,NEMO2-Se}.

\subsection{Experimental results}

The NEMO-2 detector's experimental data from measurements with Cd,
Se and Zr foils were used to estimate limits on non-Paulian transitions 
in the $^{12}$C of the plastic scintilators.
The total mass of $^{12}$C under study was 170 kg.

Fig.~3 shows non-Paulian transitions in $^{12}$C.
In Fig.~3a, 
the transition of a nucleon from the $p$-shell to the fully 
occupied $1s_{1/2}$-shell is shown. 
This process is accompanied by $\gamma$-quantum
emission, where its energy equals the energy difference between the $p$ and $s$ levels
($\sim 20$ MeV) \cite{Logan79}.
In subsequent figures, (Fig.~3b,c), the $\beta^{\pm}$ transitions of $^{12}$C 
to non-Paulian
$^{12}\tilde{\mathrm B}$ and $^{12}\tilde{\mathrm N}$
are shown when a nucleon  
falls from the $p$-shell to the fully occupied $1s_{1/2}$-shell. 
The emitted $\beta^{+}$ or
$\beta^{-}$
are distributed as ordinary $\beta$-decay spectra with an endpoint energy 
of $20$ MeV \cite{Kekez90}.              
Cuts were used for
extracting fine limits from the experiment. 

\begin{figure*}
\resizebox{0.75\textwidth}{!}{\includegraphics{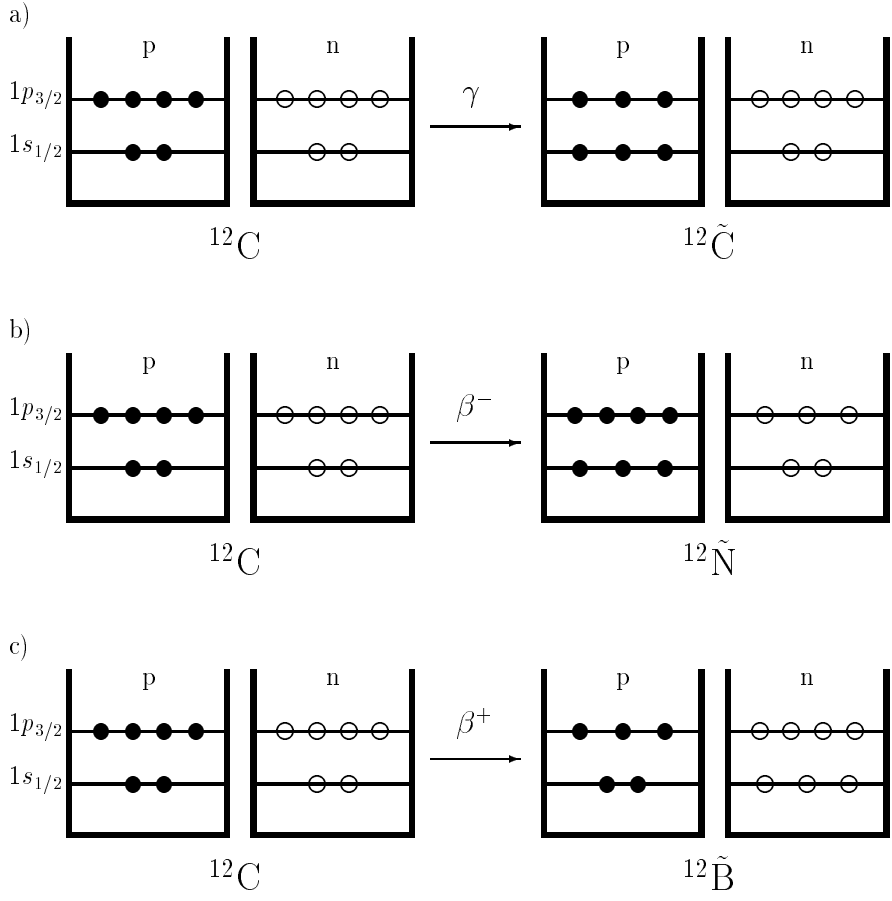}}
\label{scheme}
\caption{Schemes of non-Paulian transitions in $^{12}$C. 
(a) transition of a proton from the $p$-shell to the fully occupied
$s$-shell (a similar figure can be constructed for neutrons) 
(b) non-Paulian $\beta^{-}$ transition of 
$^{12}$C to $^{12}\tilde{\mathrm N}$;
(c) non-Paulian $\beta^{+}$ transition 
of $^{12}$C to $^{12}\tilde{\mathrm B}$.}
\end{figure*}

\subsubsection{Non-Paulian processes with high energy $\gamma$-quantum 
emission.}

High energy $\gamma$-quanta produced in a scintillator
from the non-Paulian transition to 
$^{12}\tilde{\mathrm C}$ were considered. The $\gamma$-quanta 
cross the tracking volume, 
interact with a source foil and give two
tracks and two fired scintillators. 
In the energy region $E_{\gamma}\sim 20$ MeV, pair creation probability
 in the source foil 
(45-50 mg/cm$^2$), is higher by 2-3 orders of magnitude than those for double
 Compton interactions or M\"oller scattering of Compton electrons
in the foil. The NEMO-2 detector 
was not designed to distinguish between $e^{+}$ and $e^{-}$ tracks, 
thus pairs were detected as two electron events (2e). 
A time-of-flight analysis was used to
select high energy 2e events in both simulation and experimental data.
The simulated data studied $3.8\cdot10^6$ events with  
initial $\gamma$-quanta emitted
from the scintillators.
The maximum in the summed electron energy ($E_{\rm 2e}$) spectrum 
(Fig.~4) is at the energy of the $\gamma$-quanta minus the two electron's masses.
%$E_{ee}=E^{init}_{\gamma}-2m_{e}=19$ MeV.

\begin{figure*}
\resizebox{0.75\textwidth}{!}{\includegraphics{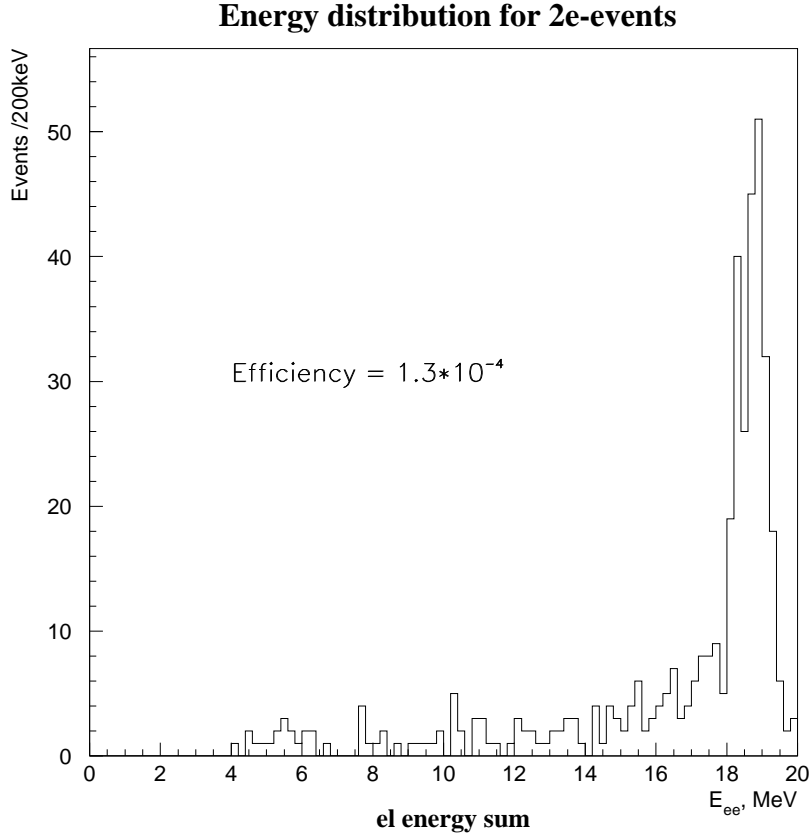}}
\label{scheme}
\caption{Simulated summed electron energy spectrum of two-electron 
events coming from the source foil for which the event was generated by a 
20 MeV $\gamma$-quanta in the plastic scintillators. A low energy cut is
applied at 4 MeV. (From Ref. \cite{ARN99}).}
\end{figure*}

No events with two tracks and summed energy 
$\ge 4$ MeV 
were found in the experiment with Se and Zr
given an exposure of 10357 h \cite{NEMO2-Se,NEMO2-Zr} and 
in the enriched Cd measurement 
with an exposure of 6588 h \cite{NEMO2-Cd}. 
The detection efficiency for
$E_{\rm 2e}\ge 4$ MeV and $\cos(\theta_{\rm 2e})>0$ is equal to $0.013\%$
for the Se and Zr sources.
In the case of the enriched Cd the efficiency should be scaled by a factor
of 0.57 because the enriched Cd is another material with a different thickness
and occupied only a half of the source plane.
From the data one can obtain a limit on the PEP-violated
transition of $^{12}$C nucleus to $^{12}\tilde{\mathrm C}$ at the 90\% C.L. of:
$$
T_{1/2} > 5.3 \cdot 10^{23} y.
$$

Also the limit
on PEP violating transitions of nucleons from the $p$-shell to the 
fully occupied
$1s_{1/2}$-shell in $^{12}$C at the 90\% C.L. is:
$$
T_{1/2} > 4.2 \cdot 10^{24} y.
$$

\subsubsection{$\beta^{\pm}$ decays to non-Paulian states.}

The search for $\beta^{\pm}$ decay processes was performed through the 
selection of two track events.
Cuts for these events require an electron to appear 
in a plastic scintillator, cross the tracking volume
and source plane then enter a plastic scintillator  on the opposite side of
the NEMO-2 detector.
Simulations of $\beta^{+}$ and $\beta^{-}$ decays of $^{12}$C to
non-Paulian states of the daughter nuclei in plastic scintillators 
was examined.
The simulated spectra are presented in Fig.~5. 
Evident here is that the efficiency for $\beta^+$ decays is lower than
$\beta^-$ decays because the detection of at least one annihilation  $\gamma$-quantum
 (511 keV) leads to the rejection of such an event. 

\begin{figure*}
\resizebox{0.75\textwidth}{!}{\includegraphics{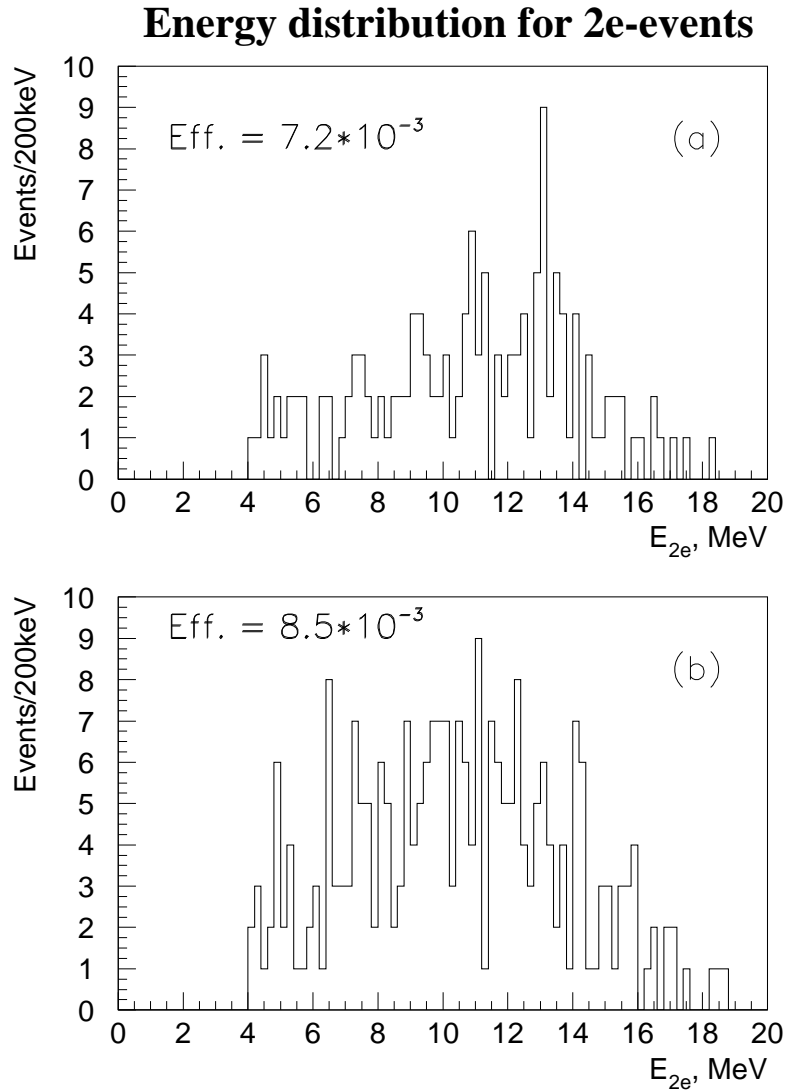}}
\label{scheme}
\caption{Energy spectra of simulated two track events: 
(a) for $\beta^+$ decay
and (b) for 
$\beta^-$ decay of $^{12}$C to non-Paulian states of daughter nuclei
 with a $20$ MeV endpoint energy. 
Here again there is a cut at 4 MeV for comparison with the experimental data.
(From Ref. \cite{ARN99}).}
\end{figure*}

The main background in the energy range up to 8 MeV 
is due to neutrons from natural sources.  
Consequently, the data used here was 
obtained in a run with a paraffin shield (2162.8 h), which
efficiently suppressed the neutron background.
Only one event with a summed energy deposit of $E > 4$ MeV 
was found in the experiment involving Se and Zr samples.
The detection efficiency of $\beta^{-}$ is equal to 0.85\% , 
and for $\beta^{+}$ is 0.72\%. As a result, one can obtain
limits on $\beta^\pm$ decays of $^{12}$C to non-Paulian states 
at the 90\% C.L.:
$$T_{1/2}>3.1\cdot 10^{24} y\;\;\;\;\;\;{\mathrm{for}}\;\;\beta^{-} $$
and
$$T_{1/2}>2.6\cdot 10^{24} y\;\;\;\;\;\;{\mathrm{for}}\;\;\beta^{+}.$$

\subsection{Conclusion and prospects for the future}

Table~\ref{table1} presents the NEMO-2 results on non-Paulian
transitions in $^{12}$C.
Due to good time-of-flight selection and a large mass of plastic
scintillator the limits for 
20 MeV gamma emission is higher by four orders of magnitude than
the previous limit \cite{Logan79}. Limits on the $\beta^{\pm}$ non-Paulian
transitions are lower than in \cite{Kekez90}, 
because of the relatively small masses involved and the low efficiency for  
crossing electron detection. Recently the limit on the transition with the emission of a 
$\gamma$-quantum was improved by BOREXINO to $T_{1/2} > 2.1\cdot 10^{27}$ y \cite{BOR04}.

\begin{table}
\caption{Limits on the non-Paulian transitions in $^{12}$C. (From Ref. \cite{ARN99}).}
\label{table1}
\begin{tabular}{llll}
\hline\noalign{\smallskip}
Channel & $\gamma$ emission & $\beta^-$ decay & $\beta^+$ decay \\
 \hline\noalign{\smallskip} 
 Window (MeV)          &  [4,20]      & [4,20] & [4,20]   \\
 Number of events      &  0           & 1       & 1    \\
 Efficiency & $1.3\cdot 10^{-4}$ & $8.5\cdot 10^{-3}$ & $7.2\cdot 10^{-3}$ \\
$T_{1/2}$  (90\% CL) present & $>4.2\cdot 10^{24}$ y & $>3.1\cdot 10^{24}$ y & 
          $>2.6\cdot 10^{24}$ y \\
\hline\noalign{\smallskip} 
$T_{1/2}$ (99.7\% CL) \cite{Logan79} & $>1.3\cdot 10^{20}$ y &  &        \\
$T_{1/2}$ (90\% CL) \cite{Kekez90}   &             & $>8\cdot 10^{27}$ y 
& $>8\cdot 10^{27}$ y \\
\noalign{\smallskip}\hline
\end{tabular}
\end{table}

The new detector, NEMO-3, which is functioning \cite{ARN05,ARN05a}, 
will improve
these limits. The amount of $^{12}$C 
is $\sim$ 40 times greater and the detection 
efficiency $\sim$ 10 times higher. Additionally 
a magnetic field is applied to distinguish $e^+e^-$ from 
$e^-e^-$ events. So expected limits which will be obtained with the NEMO-3 detector
will be three orders of magnitude higher. The sensitivity of the next generation experiment 
(SuperNEMO \cite{BAR02,OHS08}) can be estimated to be $\sim 10^{28}-10^{29}$ y 
for all processes mentioned above.

\section{Statistic of neutrinos and double beta decay \cite{BAR07}}

Do neutrinos respect the exclusion principle of its inventor?
In this paper it is assumed  that the Pauli exclusion principle is 
violated for neutrinos and therefore neutrinos obey 
(at least partly) Bose-Einstein  statistics.  
 
It may happen that due to unique properties of neutrinos, 
a violation of the Pauli principle in the  neutrino sector might 
be much stronger than in other particle sectors. Therefore 
the effects of its
violation may first be seen in neutrino physics.   

A possibility of Bose statistics for neutrinos was first 
considered in ref.~\cite{gri} where its effects on 
big bang nucleosynthesis (BBN) have been 
studied. According to \cite{gri} the change of neutrino 
statistics from pure fermionic to pure bosonic diminishes the 
primordial $^4$He abundance by $\sim 4\%$.   
%is equivalent to the 
%decrease of  number of the effective neutrino species 
%$\Delta N_{\nu} = -0.74$. 

The idea of bosonic neutrinos has been proposed independently 
in ref.~\cite{dosm}, where cosmological and astrophysical 
consequences of this hypothesis have been studied.  
Bosonic neutrinos might form a cosmological Bose condensate which 
could account for all, or a part of, the dark matter in the 
universe. ``Wrong'' statistics
of neutrinos modifies BBN, leading to the
effective number of neutrino species being smaller than three.
The conclusion in~\cite{dosm}  agrees qualitatively with the results of 
~\cite{gri}, though quantitatively a smaller decrease of 
$N_{\nu}$ is found~\cite{hansen}. 

As far as astrophysical consequences are concerned, 
dynamics of the  supernova collapse would be influenced and
spectra of a supernova neutrinos may change~\cite{dosm,kar}. 
The presence of
neutrino condensate  would enhance contributions of the Z-bursts
to the flux of the ultra high energy (UHE) cosmic rays and lead to substantial
refraction effects for neutrinos from remote sources \cite{dosm}.

A violation of the Pauli principle for neutrinos
should show up in the elementary processes where identical
neutrinos are involved. A realistic process for  this test  
is the two-neutrino  double beta decay. 
It was shown in~\cite{dosm} that 
the  probability of the decay, as well as the energy spectrum
and angular distribution of the electrons should be affected. 
Qualitative conclusions were that the pure bosonic neutrino is excluded,  
whereas 
%$ \sim 50\%$ 
a large fraction of the bosonic component 
in a neutrino state is still allowed by the present data. 
In this connection, a possibility of partly bosonic neutrinos 
should be considered. 

\subsection{Result of calculations}

In the case of $^{100}$Mo the decay proceeds mainly 
through the $1^+$ intermediate nucleus and  
the single state dominance (SSD) hypothesis should give  a good 
approximation. This is also confirmed by spectra measurements in the NEMO-3 
experiment~\cite{ARN04,SHI06}.  

Using the SSD approximation one can calculate the 
$2\nu\beta\beta$-decay half-life of $^{100}$Mo to the ground state  
for fermionic \cite{DKSS} and bosonic neutrinos  
\begin{equation} 
T_{1/2}^{f}(0^+_{g.s.}) = 6.8~10^{18} {\rm years}, ~~~
T_{1/2}^{b}(0^+_{g.s.}) = 8.9~10^{19}  {\rm years}, 
\end{equation}
so that the ratio of probabilities equals
\begin{equation}
r_0(0^+_{g.s.}) = 0.076.
\label{r0gs}
\end{equation}
The ratio $r_0(0^+_{g.s.})$ determines the weight with which  the bosonic
component enters the total rate and differential distribution. 

The higher intermediate levels can give some (basically unknown) 
contribution and this produces a systematic error in the analysis.  
To evaluate the effect of the higher states, one can consider 
the extreme case described by the  higher states dominance (HSD) approximation,
which allows one to factorize the nuclear matrix element and 
integration over the phase space of outgoing leptons.

%%%%%%%%ffff1%%%%%%%%%%%%%%%%%%%%%%%%%%%%%%%%%%%%%%%%%%%%%%%%%%%%%%%%%%%
\begin{figure}[tb]
\begin{center}
\includegraphics[width=10.0cm, height=10.0cm, angle=0]{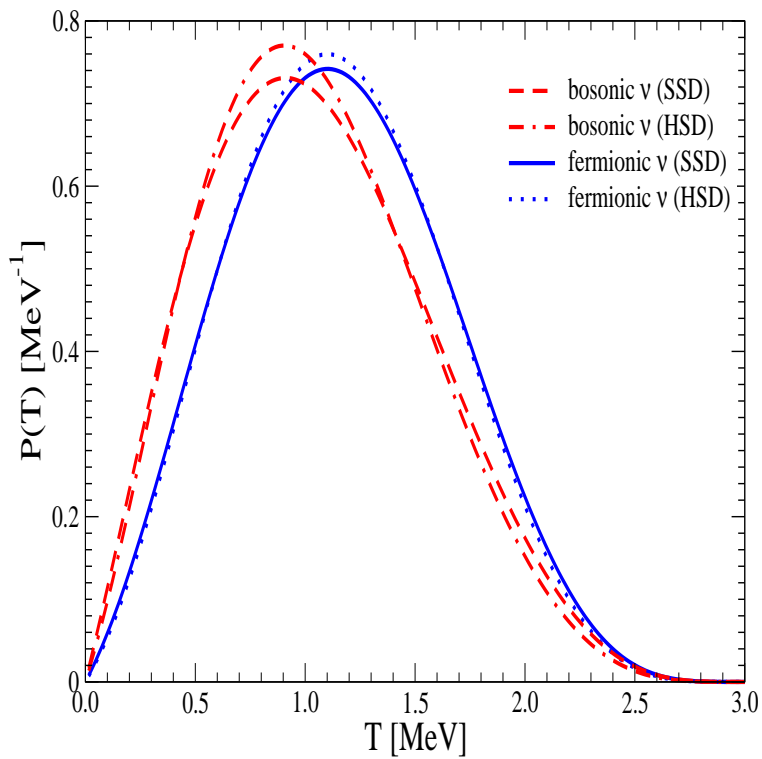}
\caption{The differential decay rates normalized to the total decay rate 
vs. the sum of the kinetic energy of outgoing electrons T for 
$2\nu\beta\beta$-decay of $^{100}$Mo to the ground state of the final nucleus.
The results are presented for the cases of pure fermionic and bosonic neutrinos.
The calculations have been performed within the single-state dominance hypothesis
(SSD) and with the assumption of dominance of higher lying states (HSD).
(From Ref. \cite{BAR07}).}
\label{mototapp}
\end{center}
\end{figure}
%%%%%%%%%%%%%%%%%%%%%%%%%%%%%%%%%%%%%%%%%%%%%%%%%%%%%%%%%%%%%%%%

%%%%%%%%ffff2%%%%%%%%%%%%%%%%%%%%%%%%%%%%%%%%%%%%%%%%%%%%%%%%%%%%%%%%%%%
\begin{figure}[tb]
\begin{center}
\includegraphics[width=10.0cm, height=10.0cm, angle=0]{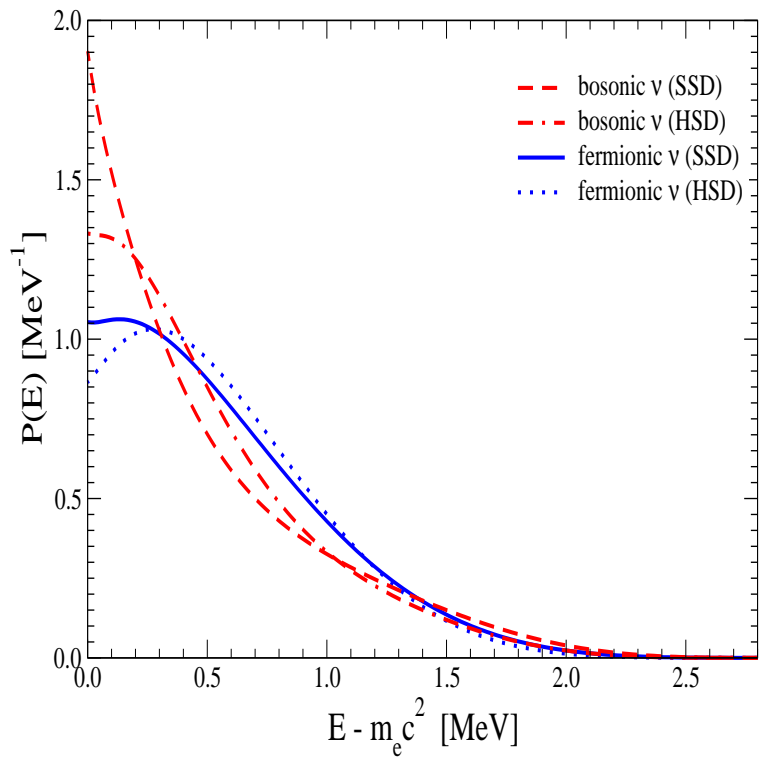}
\caption{The single electron differential decay rate normalized 
to the total decay rate vs. the electron energy for 
$2\nu\beta\beta$-decay of $^{100}$Mo to the ground state of the final nucleus.
$E$ and $m_e$ represent the energy and mass of the electron.
The results are presented for the cases of pure fermionic and bosonic neutrinos.
(From Ref. \cite{BAR07}).}
\label{mosinapp}
\end{center}
\end{figure}
%%%%%%%%%%%%%%%%%%%%%%%%%%%%%%%%%%%%%%%%%%%%%%%%%%%%%%%%%%%%%%%%

The energy spectra of electrons calculated in the 
SSD and  HSD  approximations are presented 
in the Figs. \ref{mototapp} and \ref{mosinapp}. 
The SSD approximation gives a slightly wider spectra of  
two electrons for both the fermionic and bosonic neutrinos.  
The spectra for the bosonic neutrinos are softer in both approximations. 
In particular, the maxima of SSD and HSD spectra are shifted to 
low energies for bosonic neutrinos by about 15 \% with respect 
to fermionic-neutrino  spectra. This shift does not 
depend on the approximation and therefore, can be considered  as the solid 
signature of the bosonic neutrino. 
Also the energy spectrum for single electrons becomes softer in the bosonic 
case (Fig. \ref{mosinapp}).  

\begin{figure}[tb]
\begin{center}
\includegraphics[width=10.0cm, height=10.0cm, angle=0]{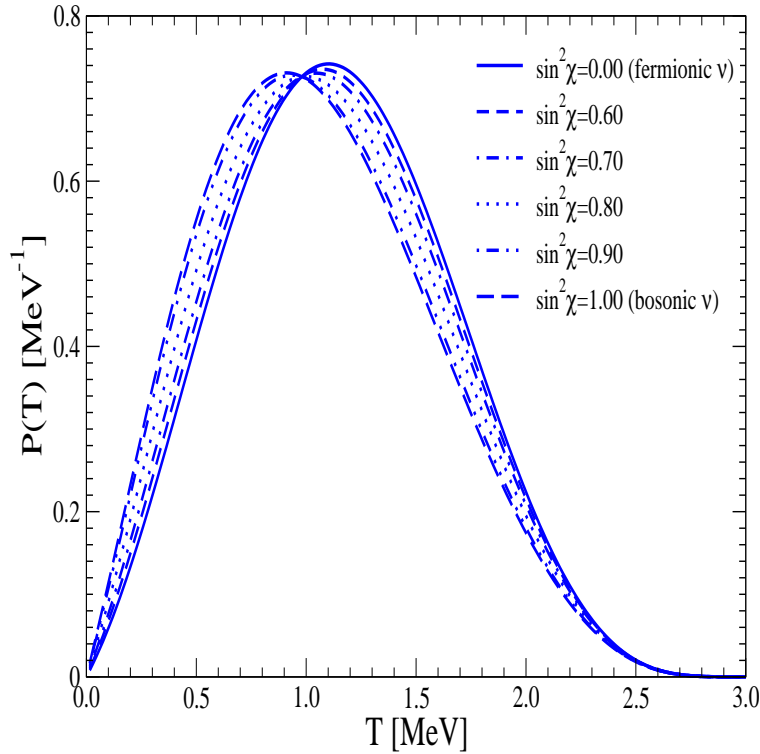}
\caption{The differential decay rates normalized to the total decay rate 
vs. the sum of the kinetic energy of outgoing electrons (T) for 
$2\nu\beta\beta$-decay of $^{100}$Mo to the ground state of the final nucleus.
The results are presented for different values of the squared
admixture of the bosonic component ($\sin^2\chi$). 
The spectra have been calculated in the SSD approximation. (From Ref. \cite{BAR07}).} 
\label{mosum}
\end{center}
\end{figure}

In Fig.~\ref{mosum} the two electron energy spectra for 
different values of the bosonic-fraction $\sin^2 \chi$ show the 
shift to smaller energies with increasing $\sin^2 \chi$. 
Due to the smallness of $r_0$, substantial shifts occur only when 
$\sin^2 \chi$ is close to 1.0. The probability of $2\nu\beta\beta$-decay 
is then equal to:
\begin{equation} 
%\be
W_{tot} = \cos^4\chi\, W_f + \sin^4\chi\, W_b  
%\nonumber\\
%&=& (1-b^2)~W_f~+~b^2~W_b,
\label{W-tot}
%\end{equation}\ee 
\end{equation}

%%%%%%%%ffff3%%%%%%%%%%%%%%%%%%%%%%%%%%%%%%%%%%%%%%%%%%%%%%%%%%%%%%%%%%%
%%%%%%%%%%%%%%%%%%%%%%%%%%%%%%%%%%%%%%%%%%%%%%%%%%%%%%%%%%%%%%%%

\begin{figure}[tb]
\begin{center}
\includegraphics[width=10.0cm, height=10.cm, angle=0]{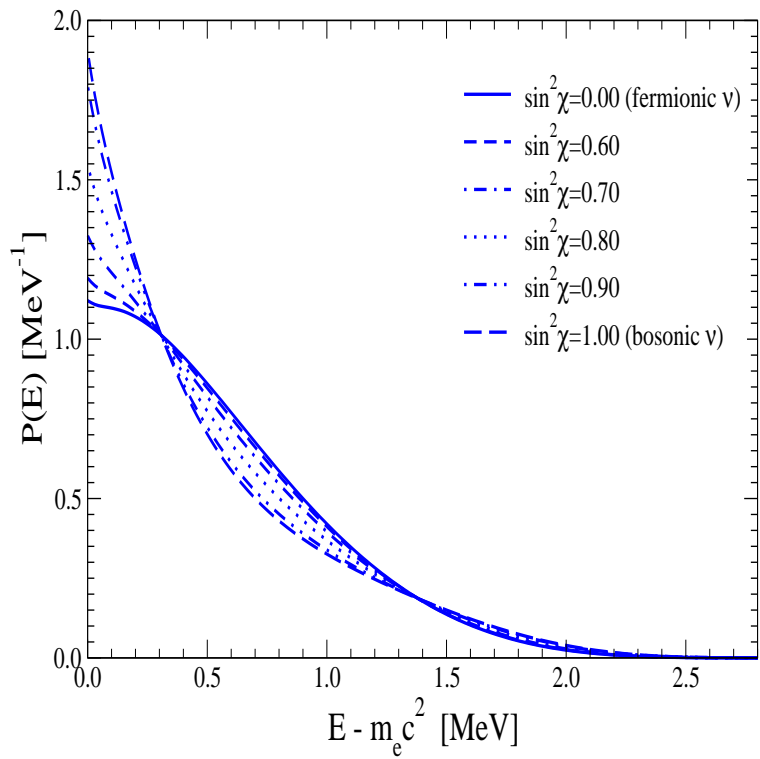}
\caption{
The single electron differential decay rate normalized 
to the total decay rate vs. the electron energy for 
$2\nu\beta\beta$-decay of $^{100}$Mo to the ground state of the final nucleus.
The results are presented for different values of the squared
admixture of the bosonic component ($\sin^2\chi$). 
The spectra have been calculated in the
SSD approximation.
The conventions are the same as in Fig. \protect\ref{mosinapp}.
(From ref. \cite{BAR07}).}
\label{mosingle}
\end{center}
\end{figure}

Fig.~\ref{mosingle} shows the energy spectra of 
single electrons for different values of  $\sin^2 \chi$. 
Note a substantial change occurs at very low energies, with 
$E = 0.3$ MeV being a fixed point.  
For $E < 0.3$ MeV the distribution increases with  $\sin^2 \chi$, 
whereas for $E =  0.3 - 1.4$ MeV it decreases.\\ 

%%%%%%%%ffff4%%%%%%%%%%%%%%%%%%%%%%%%%%%%%%%%%%%%%%%%%%%%%%%%%%%%%%%%%%%
%%%%%%%%%%%%%%%%%%%%%%%%%%%%%%%%%%%%%%%%%%%%%%%%%%%%%%%%%%%%%%%%

\subsection{Bounds on bosonic neutrinos}

Measurements of the differential characteristics of the decays
should show different shapes of the single and summed electron energies as well as
the angular distribution. 
Such information is provided  now 
by  NEMO-3 for
$^{100}$Mo, $^{82}$Se, $^{116}$Cd, $^{150}$Nd, $^{96}$Zr, $^{48}$Ca and $^{130}$Te 
\cite{ARN05a,ARN04,SHI06,baranew}. 
One should perform the statistical fit of the spectra by  
a general distribution with  $\sin^2 \chi$
being a free parameter.  
The spectral method is sensitive 
to $\sin^2\chi$ 
for nuclei and transitions with large $r_0$.  That includes  
$^{100}$Mo, as well as transitions to the excited states.

First consider the energy spectra of  
$0^+_{g.s.} \rightarrow 0^+_{g.s.}$ decay of $^{100}$Mo~\cite{ARN05a}. 
In the present paper no detailed statistical analysis 
of the spectra is applied at this time, postponing this to the time when the measurements  
are finished and there are
careful calibrations. Instead, there are some 
qualitative estimates. Here there is reasonable agreement with 
the predicted energy spectrum of the two 
electrons and the experimental points. 
Therefore one can certainly exclude the 
pure bosonic case ($\sin^2\chi  = 1$). 
Furthermore,  comparing in Fig.~\ref{mosum}
the relative shift of the bosonic maximum 
with the experimental spectrum  one can put the conservative bound on  
$\sin^2 \chi < 0.6$.
In fact,  there is no ideal agreement between the data and theoretical 
spectrum.  A better fit can be obtained for $\sin^2\chi$ between 0.4 and 0.5.  

Next a comment is made on the   
single-electron energy spectrum from $^{100}$Mo decay. 
The data agrees well with the predictions from the fermionic 
SSD mechanism, but  
some difference exists between the data 
and the fermionic HSD mechanism predictions. From this 
it was concluded that the  SSD mechanism is better 
here \cite{ARN04,SHI06}. Comparing the experimental 
data and spectra for partly bosonic neutrinos 
(Fig.~\ref{mosingle}) one  obtains $\sin^2\chi < 0.7$.  

Notice that the SSD spectrum does not show an
ideal agreement with the experimental data either. 
There is a discrepancy in the low
energy region ($E = 0.2-0.4$ MeV). 
That could be explained by the effect of 
partly bosonic neutrinos with $\sin^2 \chi \sim$ 0.5 - 0.6.\\

The full analysis 
%%(using maximal likelihood methods) 
of existing NEMO-3 information (energy and angular distributions) 
using maximal likelihood methods, 
will have a higher sensitivity to $\sin^2 \chi$. However, 
it is difficult to expect a better bound than 
 $\sin^2 \chi \sim 0.4-0.5$, because of the
existing disagreement between 
the data and Monte Carlo (MC) simulations. In fact,  
it can be just some systematic effect connected 
to the  present poor understanding of the response function of 
the detector. If in the future 
the NEMO experimental data turns out to be  in better agreement with 
the MC-simulated spectrum, 
the sensitivity to the partly bosonic neutrino will be  improved 
down to $\sin^2 \chi \sim 0.2 - 0.3$.\\ 
%% provided that we will not see the positive effect.\\

Finally the determination of the ratios of half-lives to the excited and ground state is,
\begin{equation}
r^*_{f,b} (J^\pi) \equiv 
\frac{T^{f,b}_{1/2}(J^\pi)} 
{T^{f,b}_{1/2}(0^+_{g.s.})},
\label{ratio-ex}
\end{equation}
for the fermionic and bosonic neutrinos.
For $2\nu\beta\beta$-decay of  $^{100}$Mo the ratio can be calculated 
rather reliably using 
the SSD approximation.  The advantage of this quantity
is that the EC amplitude, 
[(A,Z) $\rightarrow$ (A,Z+1) transition], which is not well determined, 
cancels in  the ratio (\ref{ratio-ex}). 

For $^{100}$Mo  the transitions to the ground ($0^+_{g.s.}$) 
and excited ($0^+_1$) states 
have been detected,  and a discrepancy has been observed.  
The corresponding experimental ratio $r^*$ equals 
\begin{equation}
r^*_{exp.} (0^+_1) \simeq 80 
\end{equation}
(NEMO-3 results \cite{ARN05a,ARN07}),  
whereas within the SSD approach the  calculated ones are
\begin{eqnarray}
r^* (0^+_1) &\simeq& 61 ~~~~~~~({\rm fermionic}~\nu) \nonumber\\
&\simeq& 73 ~~~~~~~({\rm bosonic}~\nu).
\end{eqnarray}
The bosonic neutrino fits the data slightly better but the differences are 
probably beyond the accuracy of the SSD assumption. Nevertheless, it is also 
possible to  improve the statistics in the measurements of the transition to the
excited $0^+_1$ state.

Contrary to the case of the $0^+$ excited state, the ratio 
of $2\nu\beta\beta$-decay 
half-lives to the excited $2^+$ and ground state  is expected to be 
strongly different for the bosonic and fermionic neutrinos. 
Using the SSD approximation for the 
$2\nu\beta\beta$-decay of $^{100}$Mo these are  
\begin{eqnarray}
r^* (2^+_1) &\simeq& 2.5~10^{4} ~~~~~~~({\rm fermionic}~\nu) \nonumber\\
&\simeq& 2.7~10^{2} ~~~~~~~({\rm bosonic}~\nu).
\end{eqnarray}
The $2\nu\beta\beta$-decay of $^{100}$Mo to the excited $2^+_1$ state has not been
measured yet. Using the best experimental limit on the half-life
found in \cite{BAR95} one gets
\begin{equation}
r^*_{exp} (2^+_1) > 2.2~10^{2}. 
\end{equation}
This bound is close to the bosonic prediction. Further experimental 
work in measuring this nuclear transition will allow one to analyze 
the case of the partially bosonic neutrino.  
 
\section{Conclusion}

A search was made for anomalous carbon atoms ($^{12}\tilde{\mathrm C}$), with 
three K-shell electrons. A limit on the existence of such atoms 
was determined that is $^{12}\tilde{\mathrm C}$/$^{12}$C $< 2.5\times10^{-12}$. This 
corresponds to a lifetime limit with respect to the violation of the Pauli 
principle by electrons in a carbon atom of $\tau > 2\times10^{21}$ y.

PEP was tested with the NEMO-2 detector. In the future using NEMO-3 and SuperNEMO  
the sensitivity can be increased to $\sim$ $10^{28}-10^{29}$ y.

This was the first time PEP was checked for neutrinos. Here pure bosonic neutrinos are excluded by 
the present $\beta\beta$ decay data. In the case of partly bosonic neutrinos 
the analysis of the existing data allows one to 
put the upper bound on $sin^2\chi$ of $< 0.6$. The sensitivity can be improved up 
to $sin^2\chi \sim$ 0.1-0.2.
  
\section{Acknowledgments}

%\begin{acknowledgements}
I am very thankful to Prof. S. Sutton for his useful remarks. 
Portions of this work were supported by a grant from RFBR 
(no 06-02-72553). 
This work was supported by Russian Federal Agency for Atomic Energy.
%\end{acknowledgements}

\end{document}